\documentclass[12pt,thmsb]{article}
\usepackage{sw20bams}

\input tcilatex
\QQQ{Language}{
American English
}

\begin{document}

\title{Incompressible Fields in Riemannian Manifolds}
\author{C. Viazminsky \\
IITAP, Iowa State University, Ames, Iowa 50011 \and and Department of
Physics, University of Aleppo, Syria}
\maketitle

\begin{abstract}
Incompressible fields are of special importance in electrodynamics, fluid
mechanics, and quantum mechanics. We shall derive a few expressions for such
fields in a Riemannian manifold, and show how to generate an incompressible
field from an arbitrary set of scalar differentiable functions. The concept
of compressibility removing factors of an arbitrary vector field is
introduced and utilized to obtain from an arbitrary vector field an
incompressible one that has the same vector surfaces as the original field.
A general expression for compressibility removing factors of a vector field
is derived. The method is applied to central fields.
\end{abstract}

\section{Introduction}

The equation which expresses that a vector field $L$ has a zero divergence, $%
divL=0$, appears in a number of areas$\;$of physics such as electrodynamics,
fluid mechanics, and quantum mechanics. In fact two of Maxwell's equations,
namely $div\stackrel{\rightarrow }{E}=\rho /\epsilon _0$ and $div\stackrel{%
\rightarrow }{H}=0,$ where $\stackrel{\rightarrow }{E}$ and $\stackrel{%
\rightarrow }{H}$ are the electric and magnetic fields respectively, $\rho $
is the charge density and $\epsilon _0$ is the permittivity of the medium,
reduce in the free space to the form we have mentioned. Also the continuity
equation $div\stackrel{\rightarrow }{j}+\partial \rho /\partial t=0$ \cite
{Landau} in which $\stackrel{\rightarrow }{j}$ is the current density of
some flow of matter or charge, $\rho $ is its volume density and $t$ is the
time, is reducible to this form when the volume density does not change with
time. More recently \cite{Viazminsky}, it was shown that the incompressible
vector fields in a manifold, single out a class of quantum momentum
observables that exhibit the features of the Cartesian momentum operators in
a Euclidean space. In fact it was shown that the eigenfunctions of any
member of this class is a direct generalization of De Broglie waves. The
solution of the equation $divL=0$ in the three dimensional Euclidean space
is well known. It is given in terms of a $C^2$ vector field $\stackrel{%
\rightarrow }{A}$, by $\stackrel{\rightarrow }{L}=\nabla \times \stackrel{%
\rightarrow }{A}.$ In this work

(1) We shall find a few expressions for the solutions of this equation in an 
$n$-dimensional Riemannian manifold.

(2) Show that every ,$(n-1),C^2$ functions determine an incompressible field.

(3) Introduce the concept of a compressibility removing factor of a vector
field, construct some methods for determining such factors and find the most
general expression of such factors.

(4) Show that the quotient of two compressibility removing factors of a
field is a vector surface of that field.

(5) Apply the results we have obtained to central fields and centrally
symmetric fields. \thinspace \thinspace \thinspace

We shall use tensor notations in aim to simplify the calculations and
formulae obtained. However the reader must be warned that the expressions we
get for incompressible vector fields, though valid in any coordinate system,
have no tensorial character. i.e. they do not transform as a contravariant
vector.

\section{The Field Equation}

Let $M_n\;$be a Riemannian manifold with $(x^1,....,x^n)\;$as a chart in
which the metric takes the form $ds^2=g_{ij}dx^idx^j$ , and let $%
g=det(g_{ij)}$. Let $L$ be a $C^1$ vector field on $M_n$ which has in the
given chart the form $L=\xi ^i\partial /\partial x^i.$The field $L$ is said
to be incompressible if its divergence 
\begin{equation}
divL=g^{-1/2}(g^{1/2}\xi ^i)_{,i}  \label{e1}
\end{equation}
is zero \cite{Abraham}. According to (\ref{e1}) the field $L$ is
incompressible if 
\begin{equation}
(\sqrt{g}\,\xi ^i)_{,i}=0.  \label{e2}
\end{equation}
The incompressible field equation (\ref{e2}) may be written as

\[
A_{,i}^i=o\,\,\,\,\,\,where\,\,\,\,\,A^i=\sqrt{g}\,\xi ^i. 
\]
Equation (\ref{e2}) admits obvious solutions such as 
\begin{equation}
\xi ^r=g^{-1/2},\,\,\,\,\,\xi ^j=0\,\,(j=1,....,n;j\neq r),  \label{e3}
\end{equation}
where $r$ is a fixed index chosen arbitrarily from the set $\{1,2,....,n\}$.
Since the field equation (\ref{e2}) is linear in the unknown functions $\xi
^j(or\,A^j)$ all linear combinations in the solutions (\ref{e3}) are also
solutions of the field equation. We shall see also that the commutator of
two incompressible fields is also incompressible, i.e., the set of
incompressible fields in $M_n$ forms a Lie algebra.

In the case of an one dimensional manifold $M_1$ which is coordinated by $x$
so that $ds^2=g(x)\,dx^2$, the most general solution of (\ref{e2}) is $%
L=\frac c{\sqrt{g}}\frac d{dx}$ , where $c$ is an arbitrary constant.
However $M_1$ is certainly Euclidean and the metric in $M_1$ may be reduced
to the standard form $ds^2=dX^2$, by the transformation $X=\stackunder{x_0}{%
\stackrel{x}{\int }}\sqrt{g}\,dx$. The incompressible field in the new
coordinate $X$ takes the form $L=c\;d/dX$ which is an infinitesimal motion
of $M_1.$ When $M_1$ is isometric to the real line $\Re $ or to the one
dimensional circle $S^1$, the field $L$ generates the symmetry group $%
U(X)=X+t,(t\in \Re ),$ of motions of $M_1.$

In a two-dimensional manifold $M_2$, a field $L$ is incompressible if $%
\;A_{,1}^1+A_{,2}^2=0,$ a condition which is satisfied if and only if the
ordinary differential equation $A^2dx^1-A^1dx^2=0\;$is exact; which in turn
is equivalent to the existence of a function $B$ of class $C^2$ on $M_2$
such that $B_{,1}=-A_2\;$and $B_{,2}=A_1$. It follows that the most general
form of an incompressible field in $M_2$ is of the form 
\begin{equation}
L=\frac 1{\sqrt{g}}(\frac{\partial B}{\partial x^2}\frac \partial {\partial
x^1}-\frac{\partial B}{\partial x^1}\frac \partial {\partial x^2}).
\label{e4}
\end{equation}

It is tempting to generalize the last result in a straight forward manner to
an $M_n.$ To achieve that we start with an arbitrary $C^n$ function $B$, and
take $A^i=\sqrt{g}\xi ^i$ given by 
\begin{equation}
A^1=c_1B\,_{,23...n}\,\,\,\,\,\,,\,\,A^2=c_2B_{,13....n}\;,.......\;,%
\,A^n=c_nB_{,12....n-1},  \label{e5}
\end{equation}
where $c_i$ are arbitrary constants such that $\stackunder{i=1}{\stackrel{n}{%
\sum }}c_i=0.$ We have then 
\[
A_{,i}^i=\stackrel{n}{\stackunder{i=1}{\sum }}c_i\;B_{,12....n}=0, 
\]
which proves that the field $L,$ as given by (\ref{e5}), is incompressible.
If all components of $L$ are non-zero we may satisfy $\sum c_i=0$ by 
\[
c_i=(-i)^{i-1}\tbinom{n-1}{i-1},\,\,\,\,\,\,\,\,\,\,\,\,\,\,\,\, 
\]
where no summation over $i$ is implied in the last equation. When the
dimension of the manifold is even one may choose an alternative solution $%
c_i=(-1)^{i-1}.$\ This method however, determines only a narrow class of
incompressible fields in $M_n.$ i.e., if $L=\xi ^i\partial /\partial x^i$ is
incompressible then there may not exist a function $B$ such that (\ref{e5})
is satisfied. The following theorem singles out the case in$\;$which an
incompressible field $L$ may be expressed by (\ref{e5})

\begin{theorem}
An incompressible field $L=\xi ^i\partial /\partial x^i$ can be expressed by
the form (\ref{e5}) if and only if there exist constants $c_i(i=1,...,n)$
such that $\sum c_i=0$ and 
\begin{equation}
\frac 1{c_1}A_{,1}^1=\frac 1{c_2}A_{,2}^2=....=\frac 1{c_n}A_{,n}^n.
\label{e6}
\end{equation}
If for some index $r$,$\;A_{,r}^r=0$ then we take $c_r=0$ and exclude the
corresponding ratio from the equalities (\ref{e6}).
\end{theorem}

For proof see appendix 1.

\section{General Expressions for an Incompressible Field}

\subsection{A First Expression:}

This results from taking $(n-1)$ components $\xi ^j$ of $L$ as arbitrary
functions of class $C^1$ and calculating the remaining component $\xi
^r(r\neq j)$ by the field equation (\ref{e2}); it is given by 
\begin{equation}
\sqrt{g}\xi ^r=-\stackunder{j\neq r}{\sum }\int (\sqrt{g}\xi
^j)_{,j}\;dx^r+h^r,  \label{e7}
\end{equation}
where $h^r$ is an arbitrary function of class $C^1$ independent of $x^r$.
Choosing every component $\xi ^j$ as the partial derivative with respect to $%
x^r$ of an arbitrary function $B^j$ of class $C^2,$ we write (\ref{e7} ) in
the equivalent form 
\[
\sqrt{g}\xi ^j=B_{,r}^j\,\,(j\neq r)\;,\;\;\sqrt{g}\xi ^r=-\stackunder{j\neq
r}{\sum }B_{,j}^j. 
\]
We note that the arbitrary function $h^r$ has been absorbed in $B_{,r}^j$.

\subsection{A second Expression:}

Let $e^{i_1.....i_n}$ be the totally antisymmetric unit tensor \cite{Lawden}

\ \ \ \ $e^{i_1.....i_n}=1$ if$\;(i_1....i_n)$ is an even permutation of $%
(1,2,...,n)$

\ $=-1$ if $(i_1....i_n)$ is an odd permutation of $(1,2,...,n)$

\ $=0$ if two indices are equal,

and let $B_{i_1....i_{n-2}}$, where $i_1,....,i_{n-2}$ $\in \{1,...,n\}$ and 
$i_1<i_2<....<i_{n-2}$ , be arbitrary $\tbinom n2$ functions of class $C^2$
on $M_n$.

\begin{theorem}
The field 
\begin{equation}
L=\frac 1{\sqrt{g}}e^{i_1.....i_n}B_{i_1....i_{n-2},i_{n-1}}\partial
/\partial x^{i_n}  \label{e8}
\end{equation}
\end{theorem}

is incompressible.

Proof: See appendix A2

\subsection{A Third Expression:}

Let 
\begin{equation}
\phi _k(x^1,....,x^n)\;\;\;\;\;\;\;k\in [1,n-1]  \label{e9}
\end{equation}

be $(n-1)$ arbitrary $C^2$ functions on $M_n$ that are functionally
independent. i.e. 
\[
rank(\partial \phi ^k/\partial x^j)=n-1 
\]

\begin{theorem}
The field 
\begin{equation}
L=\frac 1{\sqrt{g}}\,e^{i_1....i_n}\phi _{1,i_1}\phi _{2,i_2}.....\phi
_{n-1,i-1}\frac \partial {\partial x^{i_n}}  \label{e10}
\end{equation}
in $M_n$ is incompressible.

Proof: See appendix A3.
\end{theorem}

\begin{remark}
The expression (\ref{e10}) may be written as 
\begin{equation}
L=\frac 1{\sqrt{g}}\left| 
\begin{tabular}{llll}
$\phi _{1,1}$ & $\phi _{1,2}$ & ....... & $\phi _{1,n}$ \\ 
&  &  &  \\ 
$\phi _{n-1,1}$ & $\phi _{n-1,2}$ & ....... & $\phi _{n-1,n}$ \\ 
$\partial /\partial x^1$ & $\partial /\partial x^2$ & ....... & $\partial
/\partial x^n$%
\end{tabular}
\right|  \label{e11}
\end{equation}
\end{remark}

where the determinant must be expanded in a way that the minor of $\partial
/\partial x^i$ is the component of $\xi ^i.$

\begin{remark}
Had the functions (\ref{e9}) were functionally dependent then 
\[
rank(\partial \phi _k/\partial x^i)<n-1,\;\;\;k\in [1,n-1],\;\;i\in [1,n], 
\]
and the determinant of every $(n-1)\times (n-1)$ sub-matrix of the last
matrix vanishes. However these determinants, apart from a possible
difference in sign, are precisely the components of $L$. Hence $L$ vanishes.
\end{remark}

\begin{theorem}
Each of the surfaces 
\begin{equation}
\phi _k(x^1,....,x^n)=c_k\;\;\;\;k\in [1,n-1]  \label{e12}
\end{equation}
is a vector surface of the field (\ref{e10}), and hence the intersection of
these surfaces are the field's lines.
\end{theorem}

\TeXButton{Proof}{\proof} Let the field $L,$ as given by the expression (\ref
{e11}), acts on $\phi _k.$ It is apparent that the determinant on the right
hand-side vanishes, on the account of its k-th and last rows are equal.

Remarks and Examples

1. We have seen that in $M_2$ there exists a unique expression of an
incompressible field, namely (\ref{e4}). Applying any of the last
expressions we get (\ref{e4})

2. When we apply the second expression to $M_3$ we get 
\begin{eqnarray*}
L &=&g^{-1/2}e^{ijk}B_{i,j}\partial /\partial x^k \\
&=&g^{-1/2}[(B_{2,3}-B_{3,2})\partial /\partial
x^1+(B_{3,1}-B_{1,3})\partial /\partial x^2+(B_{1,2}-B_{2,1})\partial
/\partial x^3]
\end{eqnarray*}
where $B_1,B_2\;$and $B_3$ are arbitrary functions of class $C^2.$ This
expression is equivalent to the familiar expression$\stackrel{\rightarrow }{%
\text{ }L}=\nabla \times \stackrel{\rightarrow }{B}$ of an incompressible
field in a rectangular coordinates in a Euclidean manifold $M_3.$

Applying the third expression in $M_3$ we get

\[
L=g^{-1/2}e^{ijk}\phi _{,i}\psi _{,j}=g^{-1/2}\left| 
\begin{tabular}{lll}
$\phi _{,1}$ & $\phi _{,2}$ & $\phi _{,3}$ \\ 
$\psi _{,1}$ & $\psi _{,2}$ & $\psi _{,3}$ \\ 
$\partial /\partial x^1$ & $\partial /\partial x^2$ & $\partial /\partial
x^3 $%
\end{tabular}
\right| . 
\]
This reduces when $M_3$ is Euclidean and the coordinates are Cartesian to $%
\stackrel{\rightarrow }{L}=\nabla \phi \times \nabla \psi .$

3. The expression (\ref{e10}) of an incompressible field $L$ is equivalent
to obtaining the system of ordinary differential equations, whose integrals $%
\phi _{,k}$ $,k\in [1,n-1]$ are given, through taking the differentials$%
\;d\phi _k=0,\;$solving for $(n-1)$ of the $dx^i,$and eventually
symmetrizing the system of ordinary differential equation that has been
obtained.

\section{Reduction of a Vector Field to an Incompressible One}

Let $K=\eta ^i\partial /\partial x^i$ be a $C^1$ vector field on $M_n$, and $%
\mu (x^1,...,x^n)$ be a $C^1$ function on $M_n.$ The function $\mu $ is said
to be a compressibility removing factor ($CRF$ for short) of the vector
field $K$, or it removes the compressibility of $K$, if the vector field $%
\mu K$ is incompressible. If $\mu $ removes the compressibility of $K$ we
say that $\mu K$ reduces the field $K$. A function $\mu $ is a $CRF$ of $K$
if and only if it satisfies the equation $div(\mu K)=0$, which is equivalent
to $\mu divK+K\mu =0.$ This equation can be deduced directly, or by using
the properties of $div$ \cite{Abraham}. The equation of $CRF$ 
\begin{equation}
K\mu =-\mu divK  \label{e15}
\end{equation}
is a partial differential equation of Lagrange type. The subsidiary system
associated with this first order linear equation \cite{Piaggio} is 
\begin{equation}
\frac{dx^1}{\eta ^1}=......=\frac{dx^n}{\eta ^n}=\frac{d\mu }{-\mu \;divK}
\label{e16}
\end{equation}
We shall denote the set of incompressible fields which reduces $K$ by $L(\mu
,K).$

\begin{theorem}
Each member of the set $L(\mu ,K)$ has the same vector surfaces of $K$.
\end{theorem}

\TeXButton{Proof}{\proof} It follows from observing that the vector surfaces
of all the fields $\mu K,$ as well as the field $K,$ are determined by the
first $n$ ratios in (\ref{e16}). Or from noting that $\mu K\phi
=0\Leftrightarrow K\phi =0.$

\begin{remark}
The last theorem is applicable to every vector field of the form $\lambda K$%
, where$\;\lambda $ is any function that is not zero. This is true whether
or not $\lambda $ reduces $K$.
\end{remark}

Let 
\begin{equation}
\phi _k(x^1,....,x^n)=c_{k\,\;\;\;\;\;\;}k\in [1,n-1]  \label{e17}
\end{equation}
be functionally independent integrals of (\ref{e16}) resulting from the
first $n$ ratios. The most general arbitrary integral associated with the
field $K$ is of the form $f(\phi _1,....,\phi _{n-1})$, where $f$ is an
arbitrary function of the $\phi _k.$ Hence every vector surface of the field 
$K$ has the form $f(\phi _1,....,\phi _{n-1})=0$

\begin{theorem}
If the $CRFs$ $\mu _1$ and $\mu _2$ of the field $K$ are linearly
independent then $\mu _1/\mu _2$ is an integral of $K$.
\end{theorem}

\TeXButton{Proof}{\proof}By the equation of $CRFs$ (\ref{e15}), $\mu _2K\mu
_1=\mu _1K\mu _2$ , and hence 
\[
K(\mu _1/\mu _2)=\mu _2^{-2}(\mu _2K\mu _1-\mu _1K\mu _2)=0 
\]
Therefore $\mu _1/\mu _2$ is an integral of the subsidiary system associated
with $K,$ and $\mu _1/\mu _2=c,$ where $c$ is an arbitrary constant, is a
vector surface of $K.$

\begin{theorem}
If $\mu $ is a $CRF$ of the field $K$ then all its $CRFs$ are given by 
\begin{equation}
M=\mu f(\phi _1,....,\phi _{n-1}),  \label{e18}
\end{equation}
where$\;f$ is a $C^1$ arbitrary function in the functions $\phi _k.$
\end{theorem}

\TeXButton{Proof}{\proof} Let $M$ be a $CRF$ of the field $K$. By theorem
(4.2) $M/\mu $ is an integral of the subsidiary system associated with the
field $K$, and hence it has the form 
\[
M/\mu =f(\phi _1,....,\phi _{n-1}), 
\]
which completes the proof.

One can verify easily that the equation of $CRFs$ (\ref{e15}) is satisfied
by $M$ given by (\ref{e18}): 
\begin{eqnarray*}
(\mu f)divK+K(\mu f) &=&\mu fdivK+fK\mu +\mu Kf \\
&=&f.(\mu divK+K\mu )+0\;(becase\;Kf=0).
\end{eqnarray*}
By (\ref{e15}) the right hand side vanishes.

It follows from theorem (\ref{e7}) that if $K$ is incompressible then every $%
CRF$ of $K$ is of the form $M=f(\phi _1,....,\phi _{n-1})$.

We have seen in section 2 that a field $L=\xi ^1\partial /\partial x^1+\xi
^2\partial /\partial x^2$ is incompressible if and only if the equation 
\begin{equation}
\sqrt{g}\xi ^2dx^1-\sqrt{g}\xi ^1dx^2=0  \label{e19}
\end{equation}
is exact. It follows that $\mu $ is a $CRF$ of $L$ if and only if $\mu $ is
an integrating factor of (\ref{e19}), i.e. if and only if $\mu \sqrt{g}$ is
an integrating factor of $\xi ^2dx^1-\xi ^1dx^2=o.$

We finally mention that, although there exists an infinite family $L(\mu ,K)$
of fields which have the same vector surfaces as $K$ does, the integral
curves of these fields are distinguishable. Indeed, the differential
equations of the integral curves of a field $\mu K,$ namely 
\[
dx^i=\mu \eta ^i\;dt, 
\]
depend on the function $\mu .$

\section{Method for Determining $CRFs$}

1. If it were possible to find an integral of the system (\ref{e16}) that
contains $\mu $ then on solving for $\mu $ we get one of the required
factors.

We shall assume in what follows that we have available $(n-1)$ integrals of
the field $K$ which we denote by $\phi _k(k=1,...,n-1).$

2. Since the integrals $\phi _k$ of $K$ are also integrals to every
incompressible field $L=\mu K,$ we may find one of the fields $L=\mu K$ say $%
L=\xi ^i\partial /\partial x^i$ using the expression (\ref{e10}) of an
incompressible field in terms of its vector surfaces. Assuming that the
component $\eta ^{r\text{ }}$ of the field $K$ is not zero, then 
\begin{equation}
\mu =\xi ^r/\eta ^r=e^{i_1...i_{n-1}r}\phi _{1,i_1}....\phi
_{n-1,i_{n-1}}/\eta ^r  \label{e21}
\end{equation}
is one of the required factors.

3. Assuming that $rank(\partial \phi _j/\partial x^k)=n-1\,$where$\;k,j\in
[i,n-1]$, we may solve (\ref{e17}) for $(x^1,....,x^{n-1})$ in the form 
\begin{equation}
x^k=x^k(c_1,...,c_{n-1},x^n)\;\;\;\;\;\;\;\;\;\;k\in [1,n-1].  \label{e22}
\end{equation}
If we could obtain the solution (\ref{e22}) we substitute in (\ref{e16}) ,
separate the variables and integrate to find 
\begin{equation}
\mu =c_ne^{-\int \frac{divL}{\eta ^n}dx^n}=c_ne^{-F(c_1,....,c_{n-1},x^n)},
\label{e23}
\end{equation}
where $x^k,k\in [1,n-1],$ are substituted for in $F=\int \frac{divL}{\eta ^n}%
dx^n$ from (\ref{e22}). Substituting for $c_1,....,c_{n-1}$ in (\ref{e23})
from (\ref{e22}) we get the $n$-th integral of (\ref{e16}) 
\[
\mu (x^1,....,x^n)=f(\phi _1,...,\phi _{n-1})e^{-F(\phi _1,.....,\phi
_{n-1},x^n)}, 
\]
which is also the general form of a $CRF$.

\section{Example - Central Fields}

Let $M_n\equiv E_3$ be the three dimensional Euclidean space with Cartesian
coordinates $\stackrel{\rightarrow }{r}=(x^1=x,x^2=y,x^3=z).$ We shall say
that the field $K=\eta ^i\partial /\partial x^i$ is central if its lines are
straight lines through the origin; and centrally symmetric if it can be
reduced to the form $K=\eta (r)\partial /\partial r$, where $r=\left| 
\stackrel{\rightarrow }{r}\right| $.

Consider the centrally symmetric field 
\begin{equation}
K=x\partial /\partial x+y\partial /\partial y+z\partial /\partial z.
\label{ee1}
\end{equation}
Since $divK=3$, the equation of $CRFs\;$is 
\begin{equation}
K\mu +3\mu =0.  \label{ee2}
\end{equation}
The subsidiary system associated with (\ref{ee2}) is 
\begin{equation}
dx/x=dy/y=dz/z=-d\mu /3\mu .  \label{ee3}
\end{equation}
From (\ref{ee3}) we deduce that each of the functions $\mu _1=x^{-3},\mu
_2=y^{-3},\mu _3=z^{-3}$ is a $CRF$ of $K.\;$It follows that $\mu _1/\mu
_2=y^3/x^3,$ $\mu _1/\mu _3=z^3/x^{3\text{ }}$are two integrals of the
subsidiary system associated with $K$, and hence every vector surface of $K$
is of the form 
\begin{equation}
f(y/x,z/x)=0.  \label{ee4}
\end{equation}
According to (\ref{e18}) the general form of a $CRF$ is 
\begin{equation}
\mu =x^{-3}f(y/x,z/x).  \label{ee5}
\end{equation}
We may choose $f\;$so that we get the $CRFs$ $\mu _4=1/xyz\;$or$\;\mu
_5=(x^2+y^2+z^2)^{3/2}.$ Every vector surface of the fields $\mu
_iK(i=1,...,5)$ is given by (\ref{ee4}), which shows that the lines of all
these fields are straight lines through the origin, and hence these fields
are central. The field $K$ in spherical coordinates $(r,\phi ,\theta )$ is
written as $K=r\partial /\partial r$ which shows that it is centrally
symmetric, and that the product of $K$ by any $\mu _i,$ except $\mu _{5\text{
}},$ is not centrally symmetric; whereas $\mu _5K=r^{-2}\partial /\partial r$
is centrally symmetric. It is interesting to notice that, apart from a
multiplicative constant, $r^{-2}\partial /\partial r$ is the only
incompressible centrally symmetric field in the Euclidean space $E_{3.}$ It
is also proved easily that the most general expression for a centrally
symmetric incompressible field in $E_{n\text{ }}$ is $r^{1-n}\partial
/\partial r$ , where it is understood of course that the space has been
coordinated with a system of spherical coordinates.

Another way for obtaining a $CRF$ of the field (\ref{ee1}) is to start with
the general integral (\ref{ee4}) and construct an incompressible field $%
L=\xi ^i\partial /\partial x^i$ by formula (\ref{e10}). The first component
of $L$ is 
\[
\xi ^1=e^{ij1}\phi _{,i}\psi _{,j}=\phi _{,3}\psi _{,2}-\phi _{,2}\psi
,_3=x^{-2}, 
\]
where $\phi =y/x,\,\psi =z/x.$ By (\ref{e21}) $\mu =\xi ^1/\eta ^1=x^{-3}$
is a $CRF$, which was obtained earlier, using a different method.

The integral curves of $K$ are given by 
\[
dx^i=x^i\;dt\;\;\;(i=1,2,3). 
\]
Integrating the last system we get $x^i=x_0^i\;e^t$, where $x^i(t=0)=x_{0%
\text{ }}^i.$ The parametric equation of the integral curves is $\stackrel{%
\rightarrow }{r}=\stackrel{\rightarrow }{r_0}e^t.$

The integral curves of the incompressible centrally symmetric field $%
r^{-2}\partial /\partial r$ are 
\begin{equation}
r=(3t+r_0^3)^{-1/3},\;\;\;\;\phi =\phi _0,\;\;\;\theta =\theta _0.
\label{ee6}
\end{equation}
Since $r\geq 0$ we have $t\geq -r_0^3/3.$ When $t\rightarrow
-r_0^3/3,\,r\rightarrow 0,\;$and when $t\rightarrow +\infty ,\;r\rightarrow
+\infty .\;$In other words every point moves under the action of the
transformation (\ref{ee6}) in a straight line $(\phi =\phi _0,\theta =\theta
_0)$ ensuing from the origin when $t\rightarrow -r_0^3/3$ to occupy $%
(r_0,\phi _0,\theta _0)$ at $t=0,$ and goes to infinity as $t\rightarrow
+\infty .$

\section{A Note on the Algebraic Structure}

We know that the set of all vector fields in $M_n$, named by $\Gamma ,$ form
a Lie algebra with respect to addition, multiplication by a scalar and the
commutator operation. If $L=\xi ^i\partial /\partial x^i$ and $L_1=\eta
^k\partial /\partial x^k$ , then

\begin{eqnarray*}
L+L_1 &=&(\xi ^i+\eta ^i)\partial /\partial x^i\,\,,\;cL=(c\xi ^i)\partial
/\partial x^i, \\
\lbrack L,L_1] &=&(\xi ^i\eta _{,i}^k-\eta ^i\xi _{,i}^k)\partial /\partial
x^k.
\end{eqnarray*}
It is clear that the postulates of a vector space are satisfied and that [ ,
] is a commutator on $\Gamma .$ Let $\Im \subset \Gamma $ be the set of all
incompressible fields in $M_n$. We shall prove that $\Im $ is a sub-algebra
of $\Gamma .$ Indeed, if $L,L_1\in \Im $ then 
\[
div(L+L_1)=divL+divL_1=0,\;\;div(cL)=c\;divL=0, 
\]
and hence $\Im $ is a subspace of the vector space $\Gamma .$ Furthermore 
\cite{Abraham} 
\[
div[L,L_1]=L\;divL_1-L_1divL, 
\]
and hence $div[L,L_1]=0,$ and $\Im $ is a sub-algebra of $\Gamma .$

Let $\Pi \subset \Gamma $ be the set of all Killing fields \cite{Eisenhart}
in $M_n$. Since the divergence of every Killing field vanishes \cite
{Eisenhart}, we have $\Pi \subset \Im .$ If not empty, $\Pi $ is a Lie
algebra \cite{Eisenhart}, and $\Pi \subset \Im \subset \Gamma .$

Appendix A1. Proof of Theorem (2.1)

Necessary conditions are obvious. To prove sufficiency we define a function 
\[
B=\frac 1{c^1}\int A_{,1}^1dx^1dx^2....dx^n 
\]
Then 
\[
c_1B_{,23...n}=A^1,\;c_2B_{,13...n}=(\int
A_{,2}^2dx^1dx^2....dx^n)_{,13...n}=A^2(by\;(6)),\;etc... 
\]
Appendix A2.Proof of Theorem (3.1)

By (\ref{e1}) 
\[
\sqrt{g}divL=e^{i_1....i_n}B_{i_1....i_{n-2},i_{n-1}i_n} 
\]
For every two values of the indices $i_{n-1}$ and $i_n$ there corresponds a
unique set of values to the indices $\{i_1,...,i_{n-2}\}$ such that $%
i_1<....<i_{n-2}$ , and hence there corresponds to given values of $i_{n-1}$
and $i_n$ in the last sum two terms which are equal in absolute value: 
\[
\in (B_{i_1.....i_{n-2},i_{n-1}i_n}-B_{i_1....i_{n-2},i_ni_{n-1}})=0 
\]
The multiplier $\in $ is equal to $+1$ or $-1$ according to the permutation $%
(i_1....i_n)$ is even or odd. It follows that the sum over all the values of 
$i_{n-1}$ and $i_n$ vanishes.

Appendix A3.Proof of Theorem (3.2)

We have 
\begin{eqnarray*}
(\sqrt{g}\xi ^{i_n})_{,i_n} &=&(e^{i_1....i_n}\phi _{1,i_1}\phi
_{2,i_2}......\phi _{n-1,i_{n-1}})_{,i_n} \\
&=&e^{i_1....i_n}(\phi _{1,i_1i_n}\phi _{2,i_2}.....\phi
_{n_{-1},i_{n-1}}+.....+\phi _{1,i_1}\phi _{2,i_2}.....\phi
_{n-1,i_{n-1}i_n}) \\
&=&(\stackunder{i_1i_n}{\sum }\phi _{1,i_1i_n}\stackunder{i_2...i_{n-2}}{%
\sum }e^{i_1....i_n}\phi _{2,i_2}.....\phi _{n-1,i_{n-1}})+....... \\
&&+(\stackunder{i_{n-1}i_n}{\sum }\phi _{n-1,i_{n-1}i_n}\stackunder{%
i_1....i_{n-2}}{\sum }e^{i_1...i_n}\phi _{1,i_1.....\phi _{n-2,i_{n-2}}})
\end{eqnarray*}
We shall prove that the content of every bracket vanishes. For arbitrary
fixed values of the indices $\{i_2....i_{n-1}\}$ the first bracket contains
two terms which are equal in absolute value and different in sign, and is
written as 
\[
\in (\phi _{1,i_1i_n}-\phi _{1,i_ni_1})(\phi _{2,i_2}\phi _{3,i_3}.....\phi
_{n-1,i_{n-1}})=0 
\]
The other brackets vanish similarly, and the theorem is proved.

\end{document}